\documentclass[10pt,aps,prd,reprint,superscriptaddress]{revtex4-2}
\usepackage{overpic}
\usepackage{ifxetex}
\ifxetex
   \usepackage{polyglossia}
   \setdefaultlanguage{english}
   \usepackage{fontspec}
\else
   \usepackage[english]{babel}
   \usepackage[utf8]{inputenc}
\fi
\usepackage{lipsum}
\usepackage{hyperref}
\usepackage{amsmath,hyperref}
\usepackage{amssymb}
\usepackage{bbold}
\usepackage{enumerate}
\usepackage{float}
\usepackage{color}
\usepackage{graphicx}
\usepackage{color}
\usepackage{dsfont}
\usepackage{mathrsfs}
\usepackage{graphicx}
\graphicspath{ {images/} }

\newcommand{\beq}{\begin{equation}}
\newcommand{\eeq}{\end{equation}}
\newcommand{\beqa}{\begin{eqnarray}}
\newcommand{\eeqa}{\end{eqnarray}}
\newcommand{\bea}{\begin{eqnarray}}
\newcommand{\eea}{\end{eqnarray}}

\newcommand{\abs}[1]{\left\lvert #1\right\rvert}

\definecolor{amarillo}{cmyk}{0,0.5,1,0}

\begin{document}

\title{(Pseudo-)Kähler-Einstein geometries}

\author{Carlos G. Boiza}
\author{Jose A.R. Cembranos}
\affiliation{Departamento de F\'\i sica Te\'orica and IPARCOS, Facultad de Ciencias F\'\i sicas,
Universidad Complutense de Madrid, Plaza de Ciencias, n 1, 28040 Madrid, Spain}

\begin{abstract} 
Solutions to vacuum Einstein field equations with cosmological constant, such as the de Sitter space and the anti-de Sitter space, are basic in different cosmological and theoretical developments. It is also well known that complex structures admit metrics of this type. The most famous example is the complex projective space endowed with the Fubini-Study metric. In this work, we perform a systematic study of Einstein complex geometries derived from a logarithmic Kähler potential. Depending on the different contribution to the argument of such logarithmic term, we shall distinguish among direct, inverted and hybrid coordinates.  They are directly related to the signature of the metric and determine the maximum domain of the complex space where the geometry can be defined.
\end{abstract}

\maketitle

\renewcommand{\tocname}{Index}


\tableofcontents

\section{Introduction}

Complex manifolds have been implemented in modern theories, mainly within string frameworks. For example, the bosonic string theory is formulated on a two-dimensional differentiable manifold which is embedded on the $d$-dimensional Minkowskian space. The action of this theory is the Polyakov action \cite{Polyakov}, and it is invariant under diffeomorphisms and Weyl transformations. When the theory is quantized via path integral, we must take into account the overcounting due to the symmetries and, for this purpose, it is possible to fix the gauge by taking the called unit gauge, which fixes the metric to be the unit one; but there is still a residual symmetry due to conformal transformations, so we are dealing with a two-dimensional conformal field theory. Two-dimensional differentiable manifolds with transition functions being conformal are isomorphic to complex manifolds of dimension one (Riemann surfaces), so the theory can be formulated on an one-dimensional complex manifold.

It is interesting to implement the concept of complex manifolds within the structure of space-time. Again, the better known example of this can be found in superstring theory where there are six extra dimensions compactified on Calabi-Yau manifolds \cite{Yau}, which are a specific kind of Kähler complex manifolds. 

In this work, we are particularly interested in the implementation of the concepts of complex differential geometry on extensible (not compactified) space-time structures. Within real manifolds, it is well known the existence of solutions of the vacuum Einstein field equations with cosmological constant. This type of manifolds are called Einstein manifolds. All constant sectional curvature manifolds are of this kind. Examples of Riemannian manifolds with constant sectional curvature are the usual Euclidean space of arbitrary dimensions with the usual Euclidean metric (which is flat), and the unit $n$-sphere with the round metric (which has positive curvature). The most important examples of pseudo-Riemannian manifolds with constant sectional curvature are the de Sitter space \cite{deSitter1,deSitter2} (of positive curvature), which serves as a simple model of an accelerated expanding universe, and the anti-de Sitter space (of negative curvature), which recently acquired relevance in string theory due to the AdS/CFT correspondence proposed by Maldacena \cite{Maldacena}.

There are also well-known examples of complex Einstein manifolds. In the case of positive-definite signatures and constant bisectional curvature, we can distinguish among the complex projective space with the Fubini-Study (FS) metric (which has positive curvature), the Euclidean complex space (flat), and the unit ball (negative curvature). All this manifolds are Kähler manifolds.

The main objective of this work is to perform a systematic classification of complex Einstein geometries derived from a logarithmic Kähler potential. We will generalize the study to non-positive-definite metrics, i.e. wi will analyze pseudo-Kähler geometries and analyze their main curvature properties.

\section{Review of basic concepts}

First of all, we will summarize the main basic concepts of complex differential geometry  in order to clarify the notations and conventions we will use throughout the work. 

\subsection{Complex manifolds}  

An $m$ dimensional differentiable manifold $M$ is a topological space which can be covered by a family of open subsets $\{U_i\}$ ($M=\cup_iU_i$), where each $U_i$ is homeomorphic to an open subset $U'_i$ of $\mathbb{R}^m$: there is a map $\psi_i:U_i\subset M\rightarrow U'_i\subset\mathbb{R}^m$ which is a homeomorphism. It is also required differentiability of the maps $\Psi_{ij}=\psi_i\circ\psi^{-1}_j$ in regions where $U_i\cap U_j\neq\emptyset$. To define a complex manifold we now require open subsets $U_i$'s to be homeomorphic to open subsets $U'_i$'s of $\mathbb{C}^m$ (there is a map $\psi_i:U_i\subset M\rightarrow U'_i\subset\mathbb{C}^m$ which is a homeomorphism for each $U_i$) and maps $\Psi_{ij}=\psi_i\circ\psi^{-1}_j$ in regions $U_i\cap U_j\neq\emptyset$ to be holomorphic \cite{Nakahara}. The dimension $m$ is now understood as the complex dimension of the manifold. Note that all complex manifolds are differentiable manifolds because holomorphicity of maps automatically implies differentiability, so an $m$ dimensional complex manifold can also be seen as a $2m$ dimensional differentiable manifold. It does not happen the same in reverse order because differentiability does not necessarily implies holomorphicity, so not all differentiable manifolds are complex ones.

Since a complex manifold is a differentiable one, all the machinery developed for differentiable ones (vectors, forms, tensors, tensor fields, ...) can be used in a complex manifold. Consider a differentiable manifold which admits a tensor field $J$ of type $(1,1)$ such that, at each point $p\in M$, it squares to minus the identity $J^2_p=-id_{T_pM}$; this is called an almost complex manifold and the associated tensor field $J$ its almost complex structure \cite{GangTian}. Define the Nijenhuis tensor field $N:TM\times TM\rightarrow TM$ as the tensor which acting on two vector fields $v,w\in TM$ gives:

\begin{equation}
   N(v,w)=[v,w]+J[Jv,w]+J[v,Jw]-[Jv,Jw].
\end{equation}
It was shown by Newlander and Nirenberg \cite{NN} that the necessary and sufficient condition for a manifold to be a complex one is the vanishing of the Nijenhuis tensor $N=0$; in this case we call $J$ the complex structure of the manifold.

\subsection{Complexifications}

Consider a differentiable manifold $M$. The set of smooth functions $\mathbb{F}(M)$ is complexified \cite{Nakahara} in the following way: take two functions $g,h\in\mathbb{F}(M)$ and form a new one $f=g+ih$, which is now a complex valued function $f:M\rightarrow\mathbb{C}$; the set of functions constructed in this way forms what is called the complexification of $\mathbb{F}(M)$, denoted by $\mathbb{F}(M)^{\mathbb{C}}$. Similarly, the tangent space of $M$ at the point $p\in M$ can be complexified: take two vectors $v,w\in T_pM$ and form a new one $u=v+iw$; the set of vectors constructed in this way form the complexified tangent space $T_pM^\mathbb{C}$ (note that this is a vector space under the well defined complex addition and complex scalar multiplication). Vector fields are complexified in a similar way. A linear operator $A$ acting on the vector space $T_pM$ is extended to act on the complexified one as $A(u)=A(v+iw)=A(v)+iA(w)$, where $u=v+iw\in T_pM^\mathbb{C}$. All tensors and tensor fields can be analogously complexified. Note that complexified vector spaces have same dimension as ordinary ones.

Since an $m$ dimensional complex manifold can be seen as a $2m$ differentiable one \cite{Nakahara}, we can consider a local coordinate basis of the tangent space at $p\in M$ given by $2m$ vectors $\{\partial/\partial{x^1},...,\partial/\partial{x^m},\partial/\partial{y^1},...,\partial/\partial{y^m}\}$ associated to the real part and the imaginary part of complex coordinates $z^{\mu}=x^{\mu}+iy^{\mu}$. Its dual space is spanned by $\{dx^1,...,dx^m,dy^1,...,dy^m\}$. In this basis, the complex structure $J$ acts on vectors as $J_p(\partial/\partial{x^{\mu}})=\partial/\partial{y^{\mu}}$, $J_p(\partial/\partial{y^{\mu}})=-\partial/\partial{x^{\mu}}$. Complexifying the tangent space, we can define a new basis given by $\partial/\partial{z^{\mu}}=(\partial/\partial{x^{\mu}}-i\partial/\partial{y^{\mu}})/2$, $\partial/\partial{\overline{z}^{\mu}}=(\partial/\partial{x^{\mu}}+i\partial/\partial{y^{\mu}})/2$. The dual basis is given by $dz^{\mu}=dx^{\mu}+idy^{\mu}$, $d\overline{z}^{\mu}=dx^{\mu}-idy^{\mu}$. In this basis, $J$ acts as $J_p(\partial/\partial{z^{\mu}})=i\partial/\partial{z^{\mu}}$, $J_p(\partial/\partial{\overline{z}^{\mu}})=-i\partial/\partial{\overline{z}^{\mu}}$.

\subsection{(Pseudo-)Kähler manifolds}\label{Kähler}

Consider a complex manifold $M$ and a Riemannian metric $g$ defined on $M$ as a differentiable manifold. Its action can be generalized to the complexified tangent space at a point $p\in M$ expressed by $T_pM^{\mathbb{C}}$ \cite{Nakahara}. With this, it is possible to calculate the components of the metric with respect to the complexified basis $g_{p,\mu\nu}=g_p(\partial/\partial{z^{\mu}},\partial/\partial{z^{\nu}}), g_{p,\mu\overline{\nu}}=g_p(\partial/\partial{z^{\mu}}\partial/\partial{\overline{z}^{\nu}}), g_{p,\overline{\mu}\nu}=g_p(\partial/\partial{\overline{z}^{\mu}},\partial/\partial{z^{\nu}}), g_{p,\overline{\mu}\overline{\nu}}=g_p(\partial/\partial{\overline{z}^{\mu}},\partial/\partial{\overline{z}^{\nu}})$. If the metric $g$ satisfies $g_p(J_pv,J_pw)=g_p(v,w)$ at each point $p\in M$, where $J$ is the complex structure and $v,w\in T_pM$, it is said to be a Hermitian 
metric \cite{GangTian} and the manifold a Hermitian manifold. For a Hermitian metric, only components  $g_{\mu\overline{\nu}},g_{\overline{\mu}\nu}$ are not null: $g_{\mu\nu}=g_{\overline{\mu}\overline{\nu}}=0$. We define the Kähler form $\omega$ of a Hermitian metric $g$ as a tensor field whose action on vectors of the tangent space at a point $p\in M$ is given by $\omega_p(v,w)=g_p(J_pv,w)$, where $v,w\in T_pM$; the Kähler form is antisymmetric, so it defines a two-form. We say a metric is a Kähler metric if its Kähler form is closed: $d\omega=0$. If a metric is Kähler, the manifold, where it is defined on, is called a Kähler manifold. Consider a complex manifold $M$ and an Hermitian metric $g$ on it; the necessary and sufficient condition for the metric $g$ to be a Kähler one is $\nabla_{lc} J=0$ \cite{GangTian}, where $J$ is the complex structure and $\nabla_{lc}$ is the Levi-Civita connection (note that $\nabla_{lc}$ is well defined since a complex manifold is a differentiable one and uniquely determined by $g$, which is Riemannian by definition). A Kähler metric can always be locally expressed as $g_{\mu\overline{\nu}}=\partial_{\mu}\partial_{\overline{\nu}}K_i$ \cite{KahlerAdvanced}, where $K_i\in\mathbb{F}(U_i)$ is called the Kähler potential and $U_i$ is an open subset of $M$, and $\partial_{\mu},\partial_{\overline{\nu}}$ are partial derivatives with respect local coordinates $z^{\mu},\overline{z}^{\nu}$, respectively. It is required $K_i$ to transform as $K_j(w,\overline{w})=K_i(z,\overline{z})+\phi_{ij}(z)+\psi_{ij}(\overline{z})$ in regions where $U_i\cap U_j\neq \emptyset$, with $\phi$ holomorphic and $\psi$ antiholomorphic, and where $w$ are complex local coordinates of $U_j$ and $z$ are complex local coordinates of $U_i$; by this way, the metric $g$ transforms properly under general coordinate transformations. Reciprocally, all metrics expressed as the double derivative $\partial\overline{\partial}$ of a Kähler potential are Kähler. Similar definitions and conclusions are obtained relaxing the positive-definite condition of the metric and considering now a pseudo-Riemannian one. In this way, we
will talk about pseudo-Kähler metrics.

Since through this work we deal with metrics which will be shown to be all Kähler or pseudo-Kähler, we will often not make distinctions between them and their associated Kähler or pseudo-Kähler forms. We will use the same notation for both and we will refer to them indistinguishable.  

\subsection{Curvature of (pseudo-)Kähler metrics}\label{Curvature}

Consider a Kähler metric $g$ on a Kähler manifold $M$. We can define a connection acting on the complexified tangent space as $\nabla_{ch}:TM^{\mathbb{C}}\times TM^{\mathbb{C}}\rightarrow TM^{\mathbb{C}}$, and being compatible with the complex structure $\nabla_{ch} J=0$ \cite{GangTian}. In local coordinates, we define the Christoffel symbols of the connection as:

\begin{equation}
    \nabla_{ch,\mu}\frac{\partial}{\partial z^{\nu}}=\Gamma^{\lambda}_{\mu\nu}\frac{\partial}{\partial z^{\lambda}}+\Gamma^{\overline{\lambda}}_{\mu\nu}\frac{\partial}{\partial \overline{z}^{\lambda}},
\end{equation}

\begin{equation}
    \nabla_{ch,\mu}\frac{\partial}{\partial\overline{ z}^{\nu}}=\Gamma^{\lambda}_{\mu\overline{\nu}}\frac{\partial}{\partial z^{\lambda}}+\Gamma^{\overline{\lambda}}_{\mu\overline{\nu}}\frac{\partial}{\partial \overline{z}^{\lambda}}.
\end{equation}
For the connection to be compatible with the complex structure $\nabla_{ch} J=0$, it is required to vanish all Christoffel symbols except $\Gamma_{\mu\nu}^{\lambda},\Gamma_{\overline{\mu}\overline{\nu}}^{\overline{\lambda}}=\overline{\Gamma_{\mu\nu}^{\lambda}}$. In addition, by requiring metric compatibility $\nabla_{ch} g=0$, they result to be totally determined by the metric:

\begin{equation}
    \Gamma_{\mu\nu}^{\lambda}=g^{\lambda\overline{\rho}}\frac{\partial g_{\nu\overline{\rho}}}{\partial z^{\mu}}.
\end{equation}
This is called the Chern connection \cite{comgeo}, and can be defined similarly on Hermitian manifolds (not necessarily Kähler manifolds), but, for Kähler manifolds, and only for Kähler manifolds \cite{onlykah}, the Chern connection is the same as the Levi-Civita one extended to act on $TM^{\mathbb{C}}\times TM^{\mathbb{C}}$.

The components of the Riemann tensor associated to the Hermitian connection \cite{kahcur1,kahcur2}, in local complex coordinates, are given by:

\begin{equation}
\label{Riemann}
    R_{\mu\overline{\nu}\rho\overline{\sigma}}=-\frac{\partial^2g_{\mu\overline{\nu}}}{\partial z^{\rho}\partial \overline{z}^{\sigma}}+g^{\lambda\overline{\eta}}\frac{\partial g_{\lambda\overline{\nu}}}{\partial z^{\rho}}\frac{\partial g_{\mu\overline{\eta}}}{\partial \overline{z}^{\sigma}},
\end{equation}
and, by taking the trace in two first components, we obtain the expression for the components of a new tensor (known as Ricci tensor) in local coordinates:

\begin{equation}
\label{Ricdet}
    Ric_{\rho\overline{\sigma}}=g^{\mu\overline{\nu}}R_{\mu\overline{\nu}\rho\overline{\sigma}}=-\frac{\partial^2}{\partial z^{\rho}\partial \overline{z}^{\sigma}}(\log \abs{g}),
\end{equation}
where $g$ is the determinant of the metric.

We say that a Kähler manifold $M$ is of constant bisectional curvature \cite{GangTian} if the components in local coordinates of the Riemann tensor associated to the Hermitian connection can be written in terms of the Kähler metric ones as $R_{\mu\overline{\nu}\rho\overline{\sigma}}=\lambda_c(g_{\mu\overline{\nu}}g_{\rho\overline{\sigma}}+g_{\mu\overline{\sigma}}g_{\rho\overline{\nu}})$, where $\lambda_c$ is a real constant. For $\lambda_c>0$, the bisectinal curvature is said to be positive, for $\lambda_c=0$, null and for $\lambda_c<0$, negative.

We say that a Kähler metric $g$ is a Kähler-Einstein metric \cite{GangTian} if the Ricci tensor obtained as the trace of the Riemann tensor associated to the Hermitian connection is proportional to the metric; in local coordinates $Ric_{\mu\overline{\nu}}=\lambda_r g_{\mu\overline{\nu}}$, where $\lambda_r$ is a real constant. 

The same concepts summarized in this subsection are straightforward generalized to pseudo-Hermitian metrics.

\subsection{Fubini-Study metric}\label{FS}

Let us consider the space $N$ given by:

\begin{equation}
    N=\{Z\in\mathbb{C}^{n+1},Z\neq0\}/\sim,
\end{equation}
where we identify $Z,W\in\mathbb{C}^{n+1}$ by $Z\sim W$ if there exist a complex number $c\neq0$ such that $W=cZ$. This is called the complex projective space and it is an example of a complex manifold of dimension $n$. This manifold is isomorphic to the quotient space given by $S^{2n+1}/U(1)$, where $S^{2n+1}$ is the unit $2n+1$ hypersphere in $\mathbb{C}^{n+1}$ and $U(1)$ is the Abelian unitary group. The unit sphere can be written in terms of complex coordinates $Z$ as:

\begin{equation}
    S^{2n+1}=\left\{Z\in\mathbb{C}^{n+1}: \sum_{i=1}^{n+1}\abs{Z_i}^2=1\right\},
\end{equation}
and the quotient space identified with the manifold $N=S^{2n+1}/U(1)$ can be completely covered by the union of sets $\cup_jU_j$, where $U_j$ is given by:

\begin{equation}
    U_j(Z_j\neq0)=\left\{z\in\mathbb{C}^{n}:\sum_{\substack{i=1\\i\neq j}}^{n+1}\abs{z_i}^2>0\right\},
\end{equation}
where we have defined new coordinates $z_i=Z_i/Z_j$. 

We can now consider the Hermitian metric $g$ on $C^{n+1}$ given by $g=\sum_{i=1}^{n+1}\abs{dZ_i}^2$ and calculate the induced one by the map $\pi:M(Z)\rightarrow N(z)$. The result is the FS metric, which is given by:

\begin{equation}
\label{fs}
    g_{fs,\mu\overline{\nu}}=\partial_{\mu}\partial_{\overline{\nu}}\log\left(1+\sum_{\substack{i=1\\i\neq j}}^{n+1}\abs{z_i}^2\right).
\end{equation}

We can write the metric $g_{fs}$ in its matrix form (assuming, for instance, $j=n+1$):

\begin{equation}
g_{fs} = \frac{1}{K_{fs}^2}
\begin{pmatrix}
K_{fs}-\abs{z_1}^2 & -\overline{z}_1z_2 & \cdots & -\overline{z}_1z_n \\
-z_1\overline{z}_2 & K_{fs}-\abs{z_2}^2 & \cdots & -\overline{z}_2z_n \\
\vdots  & \vdots  & \ddots & \vdots  \\
-z_1\overline{z}_n & -z_2\overline{z}_n & \cdots & K_{fs}-\abs{z_n}^2 
\end{pmatrix},
\end{equation}
where $K_{fs}=1+\sum_{\substack{i=1\\i\neq j}}^{n+1}\abs{z_i}^2$, and, since it is a Hermitian and positive-definite matrix, the metric is Hermitian. This metric is also a Kähler metric, since it is Hermitian and can be written as the double derivative $\partial\overline{\partial}$ of the 
potential $\log K_{fs}$. We can also calculate its determinant: $\det g_{fs}=1/K_{fs}^{n+1}$. 

The determinant allows to compute all the components of the Ricci tensor by using Eq. \eqref{Ricdet}. The calculation shows that this metric is a Kähler-Einstein metric with $\lambda_{r}=n+1$ as defined in Subsection \ref{Curvature}: $R_{\mu\overline{\nu}}=(n+1)g_{fs,\mu\overline{\nu}}$, where $n$ is the dimension of the complex manifold $N$.

The group $SU(n+1)$ generates isometries of the FS metric $g_{fs}$ and also acts transitively on the complex projective space $N$. Taking this fact into account, it is enough to show the that the bisectional curvature is constant at the point $z=0$, to prove that it is constant at any point. At $z=0$, the calculus of the components of the Riemann tensor \eqref{Riemann} simplifies because all first derivatives of this metric vanish at this point. In this way, it is easy to show that $R_{\mu\overline{\nu}\rho\overline{\sigma}}|_{z=0}=(g_{fs,\mu\overline{\nu}}g_{fs,\rho\overline{\sigma}}+g_{fs,\mu\overline{\sigma}}g_{fs,\rho\overline{\nu}})|_{z=0}$. Therefore, we have a positive constant bisectional curvature with $\lambda_c=1$ as defined in Subsection \ref{Curvature}.

\subsection{Unit ball}\label{unitballsec}

Consider the space $M$ given by:

\begin{equation}
    M^{B}=\left\{Z\in\mathbb{C}^{n+1}: -\abs{Z_0}^2+\sum_{i=1}^{n}\abs{Z_i}^2=-1\right\}.
\end{equation}
We note that $Z_0\neq0$ because $\abs{Z_0}^2=1+\sum_{i=1}^{n}\abs{Z_i}^2>0$. With this, the $n$ dimensional complex manifold given by $N^{B}=M^{B}/U(1)$ can be identified with the open unit ball in $\mathbb{C}^{n}$ \cite{diffgeom}:

\begin{equation}
    N^{B}=\left\{z\in\mathbb{C}^{n}:\sum_{i=1}^{n}\abs{z_i}^2<1\right\},
\end{equation}
where we have defined new coordinates $z_i=Z_i/Z_0$. The map $\pi^{B}:M^{B}(Z)\rightarrow N^{B}(z)$ induces, from the Hermitian metric $g=-\abs{dZ_0}^2+\sum_{i=1}^{n}\abs{dZ_i}^2$, a new metric on $N^{B}$ given by:

\begin{equation}
\label{unitball}
    g_{b,\mu\overline{\nu}}=-\partial_{\mu}\partial_{\overline{\nu}}\log\left(1-\sum_{i=1}^{n}\abs{z_i}^2\right).
\end{equation}

The matrix form of $g_b$ is given by:

\begin{equation}
g_{b} = \frac{1}{K_{b}^2}
\begin{pmatrix}
K_{b}+\abs{z_1}^2 & \overline{z}_1z_2 & \cdots & \overline{z}_1z_n \\
z_1\overline{z}_2 & K_{b}+\abs{z_2}^2 & \cdots & \overline{z}_2z_n \\
\vdots  & \vdots  & \ddots & \vdots  \\
z_1\overline{z}_n & z_2\overline{z}_n & \cdots & K_{b}+\abs{z_n}^2 
\end{pmatrix},
\end{equation}
where $K_{b}=1-\sum_{i=1}^{n}\abs{z_i}^2$.

The analysis of this new metric is similar to the FS one, but we now obtain a negative constant bisectional curvature with $\lambda_r=-(n+1)$, $\lambda_c=-1$, i.e. $R_{\mu\overline{\nu}}=-(n+1)g_{b,\mu\overline{\nu}}$, and $R_{\mu\overline{\nu}\rho\overline{\sigma}}=-(g_{b,\mu\overline{\nu}}g_{b,\rho\overline{\sigma}}+g_{b,\mu\overline{\sigma}}g_{b,\rho\overline{\nu}})$.

\section{One \textit{non-direct} coordinate}

We begin to study the simplest case in which we introduce a single \textit{non-direct} coordinate. Firstly, we will consider the introduction of an \textit{inverted} coordinate and, secondly, the introduction of an \textit{hybrid} coordinate.

\subsection{One \textit{inverted} coordinate}\label{oneinvertcoord}

As mentioned in Subsection \ref{FS}, the $n$ dimensional complex projective space is isomorphic to the quotient space between the real unit $2n+1$ hypersphere and the abelian unitary group $U(1)$. This allows us to obtain the FS metric as the induced metric of the Euclidian Hermitian metric in $\mathbb{C}^{n+1}$ dimensions on the quotient space. 

The FS metric is generated by the logarithmic Kähler potential introduced in Eq. \eqref{fs}. We will use the potential in order to classify the different coordinates. For the FS metric, all coordinates are  \textit{direct} since the square of their real and imaginary parts contribute positively to the argument of the logarithmic potential. We can study the effect of introducing an \textit{inverted} coordinate, i.e. a coordinate whose square of its real and imaginary parts contribute negatively to the logarithmic Kähler potential:

\begin{equation}
\label{oneinvertmetric}
    \mathcal{G}_{\mu\overline{\nu}}=\partial_{\mu}\partial_{\overline{\nu}}\log\left(1-\abs{z_0}^2+\sum_{\substack{i=1\\i\neq j}}^{n}\abs{z_i}^2\right).
\end{equation}

Firstly, we note that this metric can be obtained as the induced of a Hermitian one in $\mathbb{C}^{n+1}$ dimensions on a quotient space, similar to the FS case. Let as consider the space $M^{I}$ given by:

\begin{equation}
    M^{I}=\left\{Z\in\mathbb{C}^{n+1}: -\abs{Z_0}^2+\sum_{i=1}^{n}\abs{Z_i}^2=1\right\}.
\end{equation}
We note that $\exists\hspace{0.5mm}i\in[1,n]:Z_i\neq0$ because $\sum_{i=1}^{n}\abs{Z_i}^2=1+\abs{Z_0}^2>0$. We now consider the quotient space given by $N^{I}=M^{I}/U(1)$; this is an $n$ dimensional complex manifold. From the observation just made, we conclude that all $N$ can be covered by the union of sets $\cup_{j}U_{j}$, where $U_{j}$ is given by:

\begin{equation}
    U_{j}(Z_j\neq 0)=\left\{z\in\mathbb{C}^{n}:-\abs{z_0}^2+\sum_{\substack{i=1\\i\neq j}}^{n}\abs{z_i}^2>-1\right\},
\end{equation}
where we have defined new coordinates $z_0=Z_0/Z_j, z_i=Z_i/Z_j$. The map $\pi^{I}:M^{I}(Z)\rightarrow N^{I}(z)$ induces, from the Hermitian metric $g=-\abs{dZ_0}^2+\sum_{i=1}^{n}\abs{dZ_i}^2$, the metric introduced in Eq. \eqref{oneinvertmetric}.  

Writing the metric $\mathcal{G}$ in its matrix form (assuming, for instance, $j=n$):

\begin{equation}
\mathcal{G} = \frac{1}{G^2}
\begin{pmatrix}
-G-\abs{z_0}^2 & \overline{z}_0z_1 & \cdots & \overline{z}_0z_{n-1} \\
z_0\overline{z}_1 & G-\abs{z_1}^2 & \cdots & -\overline{z}_1z_{n-1} \\
\vdots  & \vdots  & \ddots & \vdots  \\
z_0\overline{z}_{n-1} & -z_1\overline{z}_{n-1} & \cdots & G-\abs{z_{n-1}}^2 
\end{pmatrix},
\end{equation}
where $G=1-\abs{z_0}^2+\sum_{\substack{i=1\\i\neq j}}^{n}\abs{z_i}^2$, we see it is a pseudo-Hermitian metric since this is a Hermitian matrix and we have lost the positive-definite condition having introduced one \textit{inverted} coordinate. We can also write it as a $\partial\overline{\partial}$ derivative of the potential $\log G$, so it is a closed form ($d\mathcal{G}=0$). With this, we have found the metric $\mathcal{G}$ to be a pseudo-Kähler one. 

It is easy to compute its determinant: $\det\mathcal{G}=-1/G^{n+1}$ and the Ricci curvature in a similar way to the FS case. In fact, the components $Ric_{k\overline{l}}$ are proportional to the metric: $Ric=(n+1)\mathcal{G}$. So we conclude that $\mathcal{G}$ is a pseudo-Kähler-Einstein metric.

Finally, we will discuss the bisectional curvature. We will find it to be a constant bisectional curvature one. For this to be shown, we can use similar arguments as in FS case. The indefinite special unitary group $SU(1,n)$ is the set of matrices acting on $\mathbb{C}^{n+1}$ which preserve the Hermitian form $g(W,Z)=-\overline{W_0}Z_0+\sum_{i=1}^{n}\overline{W_i}Z_i$, where $W,Z\in\mathbb{C}^{n+1}$, with determinant one. This group acts on $N$ as an isometry and transitively, so it is enough to show the relation between Riemann curvature and metric at $z=0$. Calculus is analogous to FS one and we finally obtain the same relation $R_{i\overline{j}k\overline{l}}|_{z=0}=(\mathcal{G}_{i\overline{j}}\mathcal{G}_{k\overline{l}}+\mathcal{G}_{i\overline{l}}\mathcal{G}_{k\overline{j}})|_{z=0}$, so this metric is also of positive constant bisectional curvature with $\lambda_c=1$. We note that the pseudo-Kähler-Einstein condition could have been got directly from this last relation of positive constant bisectional curvature taking a trace in both sides, but we have used the first method of calculating the determinant because in next sections it will be important to use separate methods while both conditions are not necessarily satisfied at the same time (a metric to be Einstein does not imply to be of constant bisectional curvature). 

It is also interesting to note the analogy between the space $N^{I}$ and the real de-Sitter space. This has been emphasized by naming $N^{I}$ as the $n$ dimensional \textit{complex de-Sitter space} \cite{complexsitansit}.

\subsection{One \textit{hybrid} coordinate}\label{onehybridcoord}

In the metric $\mathcal{G}$ \eqref{oneinvertmetric}, we introduced an \textit{inverted} coordinate, so its contribution to the argument of the logarithmic potential was $-\abs{z}^2=-\mathfrak{Re}(z)^2-\mathfrak{Im}(z)^2$, where both parts (real and imaginary) contributed negatively. In this subsection, we shall consider a mixed contribution. We only reverse the sign of one of the parts (the imaginary one, for example). Therefore, we introduce an \textit{hybrid} coordinate whose contribution is given by $(z^2+\overline{z}^2)/2=\mathfrak{Re}(z)^2-\mathfrak{Im}(z)^2$. The new metric can be written as:

\begin{equation}
\label{onehybdir}
\mathcal{H}_{\mu\overline{\nu}}=\partial_{\mu}\partial_{\overline{\nu}}\log\left[1+\frac{1}{2}(z_1^2+\overline{z}_1^2)+\sum_{i=2}^{n}\abs{z_i}^2\right].
\end{equation}

The corresponding matrix form is:

\begin{equation}
\label{onehybdirmat}
\mathcal{H} = \frac{1}{H^2}
\begin{pmatrix}
-\abs{z_1}^2 & -z_1z_2 & \cdots & -z_1z_n \\
-\overline{z_1z_2} & H-\abs{z_2}^2 & \cdots & -\overline{z}_2z_n \\
\vdots  & \vdots  & \ddots & \vdots  \\
-\overline{z_1z_n} & -z_2\overline{z}_n & \cdots & H-\abs{z_n}^2 
\end{pmatrix},
\end{equation}
where $H=1+\frac{1}{2}(z_1^2+\overline{z}_1^2)+\sum_{i=2}^{n}\abs{z_i}^2$. Again, we can see that $\mathcal{H}$ is a pseudo-Kähler metric. It is pseudo-Hermitian since its matrix form is Hermitian but the positive-definite condition is not verified by the introduction of the $\textit{hybrid}$ coordinate. It is a closed form ($d\mathcal{H}$=0) because we can write it as the $\partial\overline{\partial}$ derivative of the potential $\log H$.

The determinant of this metric is also straightforward: $\det\mathcal{H}=-\abs{z_1}^2/H^{n+1}$. And it can be used to compute the Ricci tensor by taking into account Eq. \eqref{Ricdet}. We conclude that the Ricci form is proportional to the metric: $\lambda_r=n+1$: $Ric=(n+1)\mathcal{H}$, so this metric is also pseudo-Kähler-Einstein type. However, this metric does not have associated a constant bisectional curvature. In order to prove it, it is enough to compute the Riemann tensor in one given subspace where
the constant bisectional curvature condition is not verified. This is particularly simple in the subspace given by $z_{i}=0$ $\forall i\in[2,n]$, for the components with $\mu\in[2,n]$. 

\section{Generalization to an arbitrary number of \textit{non-direct} coordinates}

In this section we shall study the most general geometries that can be built from a
Käler logarithmic potential $F$ with an arbitrary number of \textit{direct}, \textit{inverted} and \textit{hybrid} coordinates.

\subsection{\textit{Hybrid} coordinates condition}\label{hybridcond}

Let us consider a form given by $\mathcal{F}_{\mu\overline{\nu}}=\partial_{\mu}\partial_{\overline{\nu}}\log F$, where:

\begin{equation}
\label{Hybridform}
F=1+\sum_{u=1}^{l}\frac{1}{2}(z_u^2+\overline{z}_u^2)+\sum_{v=l+1}^{l+m}\abs{z_v}^2-\sum_{w=l+m+1}^{l+m+p}\abs{z_w}^2,
\end{equation}
and $l,m,p$ sum up to $n$ general dimensions: $l+m+p=n$. From now on, we will use letters $u, v, w$ for complex coordinates $z_u, z_v, z_w$, respectively. So we can replace sub-indices $v=l+i,w=l+m+j$ with $i,j$, respectively, for $i\in[1,m],j\in[1,p]$, simplifying the notation. In the most general matrix form will appear nine distinguishable blocks. Let us call them $\textbf{1},\textbf{2},...,\textbf{9}$:

\begin{equation}
\label{genmatmetric}
\mathcal{F} = \frac{1}{F^2}
\begin{pmatrix}
\textbf{1} & \textbf{2} & \textbf{3} \\
\textbf{4} & \textbf{5} & \textbf{6} \\
\textbf{7} & \textbf{8} & \textbf{9}  \\
\end{pmatrix}.
\end{equation}
The main diagonal blocks $\textbf{1},\textbf{5},\textbf{9}$ are the non-mixing terms, in the sense that they are obtained acting on the potential with derivatives which do not mix different types ($u,v,w$) of complex coordinates. Therefore, they are three different square blocks: operators $\partial_{u}\partial_{\overline{u}}$, $\partial_{v}\partial_{\overline{v}}$, $\partial_{w}\partial_{\overline{w}}$ generate blocks $\textbf{1},\textbf{5},\textbf{9}$, respectively. Explicitly, they are given by:

\begin{equation}
\label{hybmat}
\textbf{1} =
\begin{pmatrix}
-\abs{u_1}^2 & -u_1\overline{u}_2 & \cdots & -u_1\overline{u}_l \\
-\overline{u}_1u_2 & -\abs{u_2}^2 & \cdots & -u_2\overline{u}_l \\
\vdots  & \vdots  & \ddots & \vdots  \\
-\overline{u}_1u_l & -\overline{u}_2u_l & \cdots & -\abs{u_l}^2 
\end{pmatrix},
\end{equation}

\begin{equation}
\textbf{5} =
\begin{pmatrix}
F-\abs{v_1}^2 & -\overline{v}_1v_2 & \cdots & -\overline{v}_1v_m \\
-v_1\overline{v}_2 & F-\abs{v_2}^2 & \cdots & -\overline{v}_2v_m \\
\vdots  & \vdots  & \ddots & \vdots  \\
-v_1\overline{v}_m & -v_2\overline{v}_m & \cdots & F-\abs{v_m}^2 
\end{pmatrix},
\end{equation}

\begin{equation}
\textbf{9} =
\begin{pmatrix}
-F-\abs{w_1}^2 & -\overline{w}_1w_2 & \cdots & -\overline{w}_1w_p \\
-w_1\overline{w}_2 & -F-\abs{w_2}^2 & \cdots & -\overline{w}_2w_p \\
\vdots  & \vdots  & \ddots & \vdots  \\
-w_1\overline{w}_p & -w_2\overline{w}_p & \cdots & -F-\abs{w_p}^2 
\end{pmatrix}.
\end{equation}
All other blocks mix different types of complex coordinates, so they are not generally square. Blocks $\textbf{2},\textbf{4}$ are generated by $\partial_{u}\partial_{\overline{v}},\partial_{v}\partial_{\overline{u}}$, respectively, and are given by:

\begin{equation}
\textbf{2} =
\begin{pmatrix}
-u_1v_1 & -u_1v_2 & \cdots & -u_1v_m \\
-u_2v_1 & -u_2v_2 & \cdots & -u_2v_m \\
\vdots  & \vdots  & \ddots & \vdots  \\
-u_lv_1 & -u_lv_2 & \cdots & -u_lv_m 
\end{pmatrix},
\end{equation}

\begin{equation}
\textbf{4} =
\begin{pmatrix}
-\overline{u_1v_1} & -\overline{u_2v_1} & \cdots & -\overline{u_lv_1} \\
-\overline{u_1v_2} & -\overline{u_2v_2} & \cdots & -\overline{u_lv_2} \\
\vdots  & \vdots  & \ddots & \vdots  \\
-\overline{u_1v_m} & -\overline{u_2v_m} & \cdots & -\overline{u_lv_m}
\end{pmatrix}.
\end{equation}
Block $\textbf{2}$ has $l\times m$ dimensions and block $\textbf{4}$ has $m\times l$. Blocks $\textbf{3},\textbf{7}$ are generated by $\partial_{u}\partial_{\overline{w}},\partial_{w}\partial_{\overline{u}}$, respectively, and are given by:

\begin{equation}
\textbf{3} =
\begin{pmatrix}
u_1w_1 & u_1w_2 & \cdots & u_1w_p \\
u_2w_1 & u_2w_2 & \cdots & u_2w_p \\
\vdots  & \vdots  & \ddots & \vdots  \\
u_lw_1 & u_lw_2 & \cdots & u_lw_p 
\end{pmatrix},
\end{equation}

\begin{equation}
\textbf{7} =
\begin{pmatrix}
\overline{u_1w_1} & \overline{u_2w_1} & \cdots & \overline{u_lw_1} \\
\overline{u_1w_2} & \overline{u_2w_2} & \cdots & \overline{u_lw_2} \\
\vdots  & \vdots  & \ddots & \vdots  \\
\overline{u_1w_p} & \overline{u_2w_p} & \cdots & \overline{u_lw_p}
\end{pmatrix}.
\end{equation}
Block $\textbf{3}$ has $l\times p$ dimensions and block $\textbf{7}$ has $p\times l$. Finally, blocks $\textbf{6},\textbf{8}$ are generated by $\partial_{v}\partial_{\overline{w}},\partial_{w}\partial_{\overline{v}}$, respectively, and are given by:

\begin{equation}
\textbf{6} =
\begin{pmatrix}
\overline{v}_1w_1 & \overline{v}_1w_2 & \cdots & \overline{v}_1w_p \\
\overline{v}_2w_1 & \overline{v}_2w_2 & \cdots & \overline{v}_2w_p \\
\vdots  & \vdots  & \ddots & \vdots  \\
\overline{v}_mw_1 & \overline{v}_mw_2 & \cdots & \overline{v}_mw_p 
\end{pmatrix},
\end{equation}

\begin{equation}
\textbf{8} =
\begin{pmatrix}
v_1\overline{w}_1 & v_2\overline{w}_1 & \cdots & v_m\overline{w}_1 \\
v_1\overline{w}_2 & v_2\overline{w}_2 & \cdots & v_m\overline{w}_2 \\
\vdots  & \vdots  & \ddots & \vdots  \\
v_1\overline{w}_p & v_2\overline{w}_p & \cdots & v_m\overline{w}_p 
\end{pmatrix}.
\end{equation}
Block $\textbf{6}$ has $m\times p$ dimensions and block $\textbf{8}$ has $p\times m$. Note that when we paste all blocks together, we obtain an Hermitian matrix, so the form $\mathcal{F}$ is generally pseudo-Hermitian (or Hemitian if it fulfills the positive-definite condition if 
there are not $\textit{hybrid}$ or $\textit{inverted}$ coordinates). 

By definition, a metric is a non-degenerate form, so the determinant of its matrix must not vanish. Let us see that this is satisfied for this general form $\mathcal{F}$ only if a very restrictive condition is satisfied for the number of $\textit{hybrid}$ coordinates: there must not be more than one coordinate of this kind ($l\leqslant1$). The determinant of a matrix can be calculated decomposing it in minors; if we consider more than one hybrid coordinates ($l>1$), we can always reduce the original matrix determinant to sufficiently small minors which cancel all of them. This happens because only minors which include $F$ terms (those ones which appear in the main diagonal of blocks $\textbf{5},\textbf{9}$) contribute, any other minors are null, and if we have $l>1$ we can always reduce the determinant to minors  with no $F$ terms (we are not considering the $F^{-2}$ common factor). This is a very restrictive result and forces us to consider only two cases: $l=0$ or $l=1$.

Before studying two cases separately, we can give a general expression for the determinant of $\mathcal{F}$ when $l\leqslant1$:

\begin{equation}
\label{detgen}
\det\mathcal{F}=\frac{(-1)^{l+p}\abs{u_1}^2+(-1)^{l+p}\delta_{l0}(1-\abs{u_1}^2)}{F^{n+1}},
\end{equation}
i.e. for the case without \textit{hybrid} coordinates: $\det\mathcal{F}_{l=0}=(-1)^p/F^{n+1}$,
whereas for one  \textit{hybrid} coordinate: $\det\mathcal{F}_{l=1}=(-1)^{p+1}\abs{u_1}^2/F^{n+1}$.  

\subsection{No \textit{hybrid} coordinates ($l=0$)}\label{geninvertcoord}

When we studied the introduction of one $\textit{inverted}$ coordinate in the Subsection \ref{oneinvertcoord}, we dealt with a logarithmic  argument $F$ given by Eq. \eqref{Hybridform} with $l=0,m=n-1,p=1$. We discussed it was possible to obtain the metric as the induced of an Euclidean one in $\mathbb{C}^{n+1}$. For this general case, we can proceed in an analogous way. Let us define the space $M^{(+)}$ as:

\begin{equation}
\label{nohybridgenspace}
    M^{(+)}=\left\{Z\in\mathbb{C}^{n+1}: \sum_{i=1}^{m+1}\abs{V_i}^2-\sum_{j=1}^{p}\abs{W_j}^2=1\right\},
\end{equation}
where we have written $Z\in\mathbb{C}^{n+1}$ as $Z=(V,W)$, with $V\in\mathbb{C}^{m+1}$, $W\in\mathbb{C}^p$. We note that $\exists\hspace{0.5mm}k\in[1,m+1]:Z_k=V_k\neq0$ because $\sum_{i=1}^{m+1}\abs{V_i}^2=1+\sum_{j=1}^{p}\abs{W_j}^2>0$. We can now consider the $n$ dimensional complex manifold $N$ given by the quotient space $N^{(+)}=M^{(+)}/U(1)$. This manifold can be covered by the union of sets $\cup_{k}U_{k}$, where $U_k$ is given by:

\begin{equation}
    U_{k}(V_k\neq 0)=\left\{z\in\mathbb{C}^{n}:\sum_{\substack{i=1\\i\neq k}}^{m+1}\abs{v_i}^2-\sum_{j=1}^{p}\abs{w_j}^2>-1\right\},
\end{equation}
where we have introduced new coordinates $v_i=V_i/V_k, w_j=W_j/V_k$. The map $\pi^{(+)}:M^{(+)}(V,W)\rightarrow N^{(+)}(v,w)$ induces, from the Euclidean metric $f=\sum_{i=1}^{m+1}\abs{dV_i}^2-\sum_{j=1}^{p}\abs{dW_j}^2$, the metric we are interested in.

It is easy to conclude the Einstein character of the metric by taking into account Eq. \eqref{Ricdet}.
We have again a pseudo-Kähler-Einstein metric with $\lambda_r=n+1$: $Ric=(n+1)\mathcal{F}_{l=0}$.
To show that it has also associated a constant bisectional curvature, we can use the same argument as in simpler cases: if the isometry group acts transitively on the manifold $N^{(+)}$, it suffices to show the relation at the point $(v,w)=0$. The proof of the relation in $(v,w)=0$ is similar to previous cases and the condition $R_{i\overline{j}k\overline{l}}|_{(v,w)=0}=(\mathcal{F}_{i\overline{j}}\mathcal{F}_{k\overline{l}}+\mathcal{F}_{i\overline{l}}\mathcal{F}_{k\overline{j}})|_{(v,w)=0,l=0}$ is satisfied. We only need to find the group which allows to generalize the result. Consider the set of matrices which act on $\mathbb{C}^{n+1}$ and preserve the Euclidean form 
$f(Z_1,Z_2)=\sum_{i=1}^{m+1}\overline{V_{1i}}V_{2i}-\sum_{j=1}^{p}\overline{W_{1j}}W_{2j}$, 
where we have used the separation $Z=(V,W)$. This set forms the group called the indefinite special unitary group $SU(p,m+1)$. This group acts on $N^{(+)}$ as an isometry and transitively, so the proof is completed. $\mathcal{F}$ has positive constant bisectional curvature: $R_{i\overline{j}k\overline{l}}=(\mathcal{F}_{i\overline{j}}\mathcal{F}_{k\overline{l}}+\mathcal{F}_{i\overline{l}}\mathcal{F}_{k\overline{j}})|_{l=0}$ ($\lambda_c=1$).

\subsection{One \textit{hybrid} coordinate ($l=1$)}\label{genhybridsec}

We have previously studied a metric obtained from a form $F$ with $l=1,m=n-1,p=0$ in Subsection \ref{onehybridcoord}. In this case, we shall discuss a generalize form $F$ which also contains \textit{inverted} coordinates: $l=1,m+p=n-1$.

When studying the simpler case $p=0$, the metric only contained terms as the ones contained in blocks \textbf{1}, \textbf{5}, \textbf{2}, \textbf{4}. In the general case, there are terms related to every block. However, $l=1$, so block \textbf{1} reduces to one term $\textbf{1}=-\abs{u_1}^2$, blocks \textbf{4} and \textbf{7} reduce to vectors, and blocks \textbf{2} and \textbf{3} to their conjugate transposed ones. In any case, by using Eq. \eqref{Ricdet} is easy to obtain: $Ric=(n+1)\mathcal{F}_{l=1}$. So, in it is also a pseudo-Kähler-Einstein metric in this general case. 

However, it does not have associated a constant bisectional curvature. We can consider again the subspace given by $v_i=w_j=0$ $\forall i\in[1,m],\forall j\in[1,p]$. It is enough to show that the constant bisectional curvature relation does not hold in this subspace to complete the proof. For this general case $m+p=n-1$, the same product of metrics relation as in the simpler case $p=0$ is satisfied: $(\mathcal{F}_{i\overline{j}}\mathcal{F}_{k\overline{l}}+\mathcal{F}_{i\overline{l}}\mathcal{F}_{k\overline{j}})|_{v_i=w_j=0,l=1}=(\delta_{ij}\delta_{kl}+\delta_{il}\delta_{kj})/F^2_{l=1}(v_i=w_j=0)$, where all indices are associated  with \textit{direct} coordinates $i,j,k,l\in[1,m]$. It also holds if all indices are associated with \textit{inverted} coordinates $i,j,k,l\in[1,p]$. However, when calculating the components of the Riemann tensor, the \textit{hybrid} coordinate appears explicitly in calculations trough the contraction between the metric and its first derivatives. Therefore, if we calculate components $R_{i\overline{j}k\overline{l}}$ with all indices associated to \textit{direct} coordinates $i,j,k,l\in[1,m]$, an extra term not given by the double derivative of the metric contributes. The double derivative gives a contribution equal to the product of metrics, but the contraction between the metric and its first derivatives adds a negative term of the form $-\delta_{il}\delta_{kj}/F^2_{l=1}(v_i=w_j=0)$. With this, the components of the Riemann tensor in this subspace are given by $R_{i\overline{j}k\overline{l}}|_{v_i=w_j=0}=\delta_{ij}\delta_{kl}/F^2_{l=1}(v_i=w_j=0)$ and do not satisfy the constant bisectional curvature relation. For terms of the Riemann tensor associated to \textit{inverted} coordinates $i,j,k,l\in[1,p]$ we obtain the same expression and, therefore, we can conclude that this type of metrics do not have a constant bisectional curvature.  

\section{Signatures of \textit{inverted} and \textit{hybrid} coordinates}\label{hybcon}

We have discussed how the introduction of new \textit{non-direct} coordinates to the logarithmic Käler potential \eqref{Hybridform} allows us to define new pseudo-Kähler-Einstein metrics non-positive-definite. 
In this section we would like to clarify that these new coordinates (\textit{inverted} and \textit{hybrid}) have associated opposite signatures to the \textit{direct} coordinates.

First of all, since we are dealing with metrics which are all Hermitian or pseudo-Hermitian, they can
be locally diagonalized by an unitary matrix and their eigenvalues are all real \cite{linalgebra}. This is particularly useful because the signature of the metrics is completely determined by the sign of their eigenvalues.

\subsection{One \textit{hybrid} and arbitrary number of \textit{direct} coordinates ($l=1$, $p=0$)}\label{hybcontrfs}

Consider the metric $\mathcal{H_{\mu\overline{\nu}}}$ \eqref{onehybdir} introduced in Subsection \ref{onehybridcoord} or, analogously, the metric corresponding to the form \eqref{Hybridform} with $l=1$, $p=0$. The eigenvalues $\lambda$ are solutions of the equation $\det(\mathcal{H}-\lambda\mathbb{1})=0$, where $\mathcal{H}$ is the matrix \eqref{onehybdirmat}. By subtracting the $H^{-2}$ factor, all eigenvalues must satisfy:

\begin{equation}
\label{rootdir}
(\lambda-H)^{n-2}\left[\lambda^2-\left(1+\frac{1}{2}(z_1-\overline{z}_1)^2\right)\lambda-\abs{z_1}^2H\right]=0,
\end{equation}
where $H$ and the \textit{hybrid} coordinate $z_1$ were introduced in Subsection \ref{onehybridcoord}. This equation has $n$ real roots. $n-2$ of them are $H$ and the other two are the solutions to the second degree equation between square brackets:

\begin{equation}
\lambda_{\pm}=\frac{B_A\pm\sqrt{{B_A}^2+4\abs{z_1}^2H}}{2}.
\end{equation}
Here $B_A=1+(z_1-\overline{z}_1)^2/2$. We find a positive root $\lambda_+$ and a negative one $\lambda_-$ which, together with the other ones, complete the $n$ eigenvalues. Note that $H$ must be positive for $\log H$ to be well defined. Therefore, the signature of the metric $\mathcal{H_{\mu\overline{\nu}}}$ \eqref{onehybdir} is $(-,+,+,... +)$. The minus signature corresponds to the introduction of the \textit{hybrid} coordinate, whereas the positive signatures are provided by the presence of the \textit{direct} ones.

On the other hand, the computation of the eigenvalues allow us to check the value of the determinant of the metric. Taking now into account the $H^{-2}$ factor into the multiplication of the $n$ eigenvalues, we find:
\begin{equation}
\det \mathcal{H}=\frac{H^{n-2}\lambda_+\lambda_-}{H^{2n}}=-\frac{\abs{z_1}^2}{H^{n+1}},
\end{equation}
which agrees with the result presented in Subsection \ref{onehybridcoord}.

\subsection{One \textit{hybrid} and arbitrary number of \textit{inverted} coordinates ($l=1$, $m=0$)}\label{hybcontrinv}

We now consider the metric \eqref{genmatmetric} corresponding to the logarithmic argument \eqref{Hybridform} when $l=1$, $m=0$. Again, to find the eigenvalues $\lambda$, we need to solve the equation $\det(\mathcal{\mathcal{F}}-\lambda\mathbb{1})=0$, where $\mathcal{F}$ is the matrix \eqref{genmatmetric} (we again do not take into account the multiplicative factor $F^{-2}$ of 
matrix $\mathcal{F}$ when calculating the eigenvalues). The eigenvalues equation reads:

\begin{equation}
(\lambda+F)^{n-2}\left[\lambda^2+\left(1+\frac{1}{2}(u_1+\overline{u}_1)^2\right)\lambda+\abs{u_1}^2F\right]=0,
\end{equation}
where $F$ and the \textit{hybrid} coordinate $u_1$ were introduced in Subsection \ref{hybridcond}. Therefore, $n-2$ eigenvalues are $-F$ and the other two correspond also to the roots of a second degree equation, which in this case reads:

\begin{equation}
\lambda^{(1,2)}_{-}=\frac{-B_B\pm\sqrt{{B_B}^2-4\abs{u_1}^2F}}{2},
\end{equation}
where $B_B=1+(u_1+\overline{u}_1)^2/2$. In this case, it is easy to conclude that the last two eigenvalues, $\lambda^{(1)}_{-}$, and $\lambda^{(2)}_{-}$, are also negative. Note that the manifold under study is limited by the domain where $F$ is positive. Therefore, the signature of this metric with 
$l=1$, $m=0$ is negative-definite: (-,-,-,....-). So, by changing the sign of the entire metric, we can define a positive-definite Kähler-Einstein geometry. We shall discuss this question in the next section. 

We can again calculate the determinant as the product of the eigenvalues. Taking into account the 
multiplicative $F^{-2}$ factor of matrix $\mathcal{F}$, we find:

\begin{equation}
\det \mathcal{F}=\frac{(-1)^{n-2}F^{n-2}\lambda^{(1)}_-\lambda^{(2)}_-}{F^{2n}}=(-1)^n\frac{\abs{u_1}^2}{F^{n+1}},
\end{equation}
which agrees with the result presented in Eq. \eqref{detgen} in Subsection \ref{hybridcond} with $l=1$, $m=0$.

\subsection{One \textit{hybrid}, one \textit{direct} and one \textit{inverted} coordinate ($l=1$, $m=1$, $p=1$)}\label{hybcontrgen}

For a general case, the computation of the eigenvalues is not so simple. We can illustrate the general situation by considering the metric corresponding to the argument \eqref{Hybridform} with $l=1$, $m=1$, and $p=1$. Proceeding in the same way than in previous Subsections \ref{hybcontrfs} and \ref{hybcontrinv}, the eigenvalues  $\lambda$ are determined by the equation $\det(\mathcal{F}-\lambda\mathbb{1})=0$, where $\mathcal{F}$ is the matrix \eqref{genmatmetric}. We again omit the $F^{-2}$ factor of matrix $\mathcal{F}$ when calculating the eigenvalues. Such equation reads:

\begin{equation}
\label{eqhybdicinv}
\lambda^3+A\lambda^2-CF\lambda-\abs{u_1}^2F^2=0,
\end{equation}
where $A=\abs{u_1}^2+\abs{v_1}^2+\abs{w_1}^2$, $C=1+(u^2_1+\overline{u}^2_1)/2$, and $F$ and coordinates $u_1$, $v_1$, $w_1$ were introduced in Subsection \ref{hybridcond}. The solutions of this equation are not as trivial as the ones discussed above, but we can analyze the sign of the roots using the Descartes rule. Let us note that $A$ and $F$ are positive, whereas the sign of $C$ is not completely determined. There are regions where it is positive and regions where it is negative. In spite of this, the number of sign changes in the coefficients of the equation is always one. It ensures the existence of only one real positive root. The other two roots could be two different real negative ones, one real negative root of multiplicity two or two complex roots (one complex and its conjugate). However, this last possibility is discarded because of the Hermitian character of the metric, which ensures the presence of real roots, as we noted at the beginning of this Section \ref{hybcon}. Therefore, we can conclude that the signature fo the metric is $(-,+,-)$. As in previous Cases \ref{hybcontrfs}, \ref{hybcontrinv}, the introduction of the  \textit{hybrid} coordinate $u_1$ is associated to a negative signature. In the same way, the \textit{inverted} coordinate provides a negative signature, in opposition to the \textit{direct} one. 

On the other hand, the Vieta formulas allow us to calculate again the determinant as the product of the eigenvalues. Taking into account the multiplicative $F^{-2}$ factor of matrix $\mathcal{F}$, we find:

\begin{equation}
\det \mathcal{F}=\frac{\lambda_1\lambda_2\lambda_3}{F^6}=\frac{\abs{u_1}^2}{F^4},
\end{equation}
where $\lambda_1$, $\lambda_2$, $\lambda_3$ are the solutions to the eigenvalue Eq. \eqref{eqhybdicinv}. This result agrees with the general result presented in Eq. \eqref{detgen} with $l=1$, $m=1$, $p=1$. It agrees also with the general statement claimed at the beginning of this section. \textit{Hybrid} and \textit{inverted} coordinates are associated with negative signatures, whereas \textit{direct} coordinates provide positive signatures. 

\section{Negative curvatures}\label{NegCursec}

In the previous sections (except for the Subsection \ref{unitballsec}), we have worked only with
(pseudo-)Kälher-Einstein geometries with positive curvatures.  This is due to our convention for defining the metric: $\mathcal{F}_{\mu\overline{\nu}}=\partial_{\mu}\partial_{\overline{\nu}}\log F$, in
terms of the argument $F$ given by Eq. \eqref{Hybridform}. It is easy to conclude that the definition
$\mathcal{F}^{(-)}_{\mu\overline{\nu}}=-\partial_{\mu}\partial_{\overline{\nu}}\log (2-F)$ provides a (pseudo-)Kälher-Einstein metric with negative curvature. Wrote in that way, \textit{direct} and \textit{hybrid} coordinates are associated with positive signatures, whereas \textit{inverted} coordinates are related to negative signatures. 
  
\subsection{Negative bisectional curvatures}\label{NegBisCursec}

We can illustrate this fact by analyzing cases with constant bisectional curvatures.
When discussing general metrics with no \textit{hybrid} coordinates in Subsections \ref{oneinvertcoord} and \ref{geninvertcoord}, we only found metrics with constant positive bisectional curvature despite of the appearance of \textit{inverted} coordinates. We now discuss how to construct metrics with constant negative bisectional curvatures. Consider the space $M^{(-)}$ given by:

\begin{equation}
\label{neggenspace}
    M^{(-)}=\left\{Z\in\mathbb{C}^{n+1}: \sum_{i=1}^{m}\abs{V_i}^2-\sum_{j=1}^{p+1}\abs{W_j}^2=-1\right\},
\end{equation}
where we have written $Z\in\mathbb{C}^{n+1}$ as $Z=(V,W)$, with $V\in\mathbb{C}^{m}$, 
$W\in\mathbb{C}^{p+1}$, and $m+p=n$. We have only reversed the sign of one side of the equality with respect to the definition of the space $M^{(+)}$ given by Eq. \eqref{nohybridgenspace} in the previous case with positive curvature. We also note that $\exists\hspace{0.5mm}j\in[1,p+1]:Z_j=W_j\neq0$ because $\sum_{j=1}^{p+1}\abs{W_j}^2=1+\sum_{i=1}^{m}\abs{V_i}^2>0$. We can now consider the $n$ dimensional complex manifold $N^{(-)}$ given by the quotient space 
$N^{(-)}=M^{(-)}/U(1)$. This manifold can be covered by the union of sets $\cup_{k}U_{k}$, where $U_k$ is given by:

\begin{equation}
    U_{k}(W_k\neq 0)=\left\{z\in\mathbb{C}^{n}:\sum_{i=1}^{m}\abs{v_i}^2-\sum_{\substack{j=1\\j\neq k}}^{p}\abs{w_j}^2<1\right\},
\end{equation}
where we have introduced new coordinates $v_i=V_i/W_k, w_j=W_j/W_k$. The map 
$\pi^{(-)}:M^{(-)}(V,W)\rightarrow N^{(-)}(v,w)$ induces, from the Euclidean metric $f=\sum_{i=1}^{m}\abs{dV_i}^2-\sum_{j=1}^{p+1}\abs{dW_j}^2$, a metric on $N^{(-)}$ given by:
\begin{equation}
\label{gennegcurmetric}
\mathcal{F}^{(-)}_{\mu\overline{\nu}}=-\partial_{\mu}\partial_{\overline{\nu}}\log\left(1-\sum_{i=1}^{m}\abs{v_i}^2+\sum_{\substack{j=1\\j\neq k}}^{p+1}\abs{w_j}^2\right).
\end{equation}
The analysis of this metric is similar to the analysis of the induced metric made in the previous case for the space $M^{(+)}$, defined by Eq. \eqref{nohybridgenspace}. However, in this case, the metric has a negative constant bisectional curvature with $\lambda_r=-(n+1)$, and $\lambda_c=-1$: 
$Ric=-(n+1)\mathcal{F}^{(-)}$, and $R_{i\overline{j}k\overline{l}}=
-(\mathcal{F}^{(-)}_{i\overline{j}}\mathcal{F}^{(-)}_{k\overline{l}}+
\mathcal{F}^{(-)}_{i\overline{l}}\mathcal{F}^{(-)}_{k\overline{j}})$. Note that we have only changed the definition of the space $M^{(-)}$ with respect to the space $M^{(+)}$, by reversing the sign of one side of the equality, that defines it in Eq. \eqref{nohybridgenspace}; the metric $f$ from which we induce the new one $\mathcal{F}^{(-)}$ has not been modified. However, written as in Eq. \eqref{gennegcurmetric}, the $v_i$ (\textit{direct}) coordinates have associated positive signatures, whereas the $w_j$ (\textit{inverted}) coordinates provide negative signatures. 

For instance, the particular case $p=0$ corresponds to the unit ball geometry introduced in Subsection \ref{unitballsec}. On the other hand, for the case $p=1$, the space $N^{(-)}$ may be identified as the $n$ dimensional \textit{complex anti de-Sitter space} \cite{complexsitansit}.  

\section{Conclusions}

Metrics that are solutions of vacuum Einstein field equations with cosmological constant are known as
Einstein geometries. In real structures, they have played a fundamental role not only in
theoretical developments but also in cosmological applications. In this work we have analyzed Einstein
metrics in complex manifolds by performing a systematic study of quadratic contributions to the logarithmic Kähler potential. After reviewing well-known examples as the Fubini-Study metric or
the Unit ball geometry, we have found new Kähler-Einstein and pseudo-Kähler-Einstein metrics by the introduction not only of \textit{direct} coordinates, but also \textit{inverted} and \textit{hybrid} coordinates. 

The square modulus of \textit{direct} coordinates contribute positively to the argument function $F$ of the logarithmic Kähler potential. The square modulus of \textit{inverted} coordinates contribute negatively to the same argument $F$. On the contrary, a \textit{hybrid} coordinate is characterized by
an opposite contribution of the square of its real part with respect to the contribution of the square of its imaginary part. The signature of \textit{direct} coordinates is positive, whereas the  signature of \textit{inverted} coordinates is negative. In contrast, the signature of \textit{hybrid} coordinates depend on the curvature of the metric. For positive curvatures, their signature is negative, whereas it is positive for negative curvatures.

We have found a very restrictive condition for a (pseudo-)Kähler-Einstein metric to be non-degenerate: It must have associated no more than one \textit{hybrid} coordinate. This has allowed us to separate the most general cases in two different classes: Metrics with one \textit{hybrid} coordinate and an arbitrary number of \textit{direct} and \textit{inverted} coordinates; and metrics with no \textit{hybrid} coordinates and also an arbitrary number of \textit{direct} and \textit{inverted} coordinates. There are important differences between these two types of geometries. The latter one not only corresponds to Kähler-Einstein or pseudo-Kähler-Einstein metrics, but also these geometries are characterized by a constant bisectional curvature. This is not the case of the former one, whose geometries own less isometries. In any case, we have built explicit examples of both types, where it is easy to check our general results associated with different geometrical features, such as determinants or curvatures.

\begin{acknowledgments}
This work was partially supported by the MICINN (Spain) project PID2019-107394GB-I00 (AEI/FEDER, UE).
JARC acknowledges support by Institut Pascal at Université Paris-Saclay during the Paris-Saclay Astroparticle Symposium 2021, with the support of the P2IO Laboratory of Excellence (program “Investissements d’avenir” ANR-11-IDEX-0003-01 Paris-Saclay and ANR-10-LABX-0038), the P2I axis of the Graduate School Physics of Université Paris-Saclay, as well as IJCLab, CEA, IPhT, APPEC, the IN2P3 master projet UCMN and EuCAPT.
This research was supported by the Munich Institute for Astro- and Particle Physics (MIAPP) which is funded by the Deutsche Forschungsgemeinschaft (DFG, German Research Foundation) under Germany´s Excellence Strategy – EXC-2094 – 390783311.
\end{acknowledgments}

\bibliographystyle{unsrt}
\bibliography{bibliografia}

\begin{thebibliography}{10}

\bibitem{Polyakov}
Stanley Deser and B.~Zumino.
\newblock {A Complete Action for the Spinning String}.
\newblock {\em Phys. Lett. B}, 65:369--373, 1976.

\bibitem{Yau}
Shing-Tung Yau.
\newblock {On the Ricci curvature of a compact K\"ahler manifold and the
  complex Monge-Amp\'ere equation, I}.
\newblock {\em Commun. Pure Appl. Math.}, 31(3):339--411, 1978.

\bibitem{deSitter1}
W.~de~Sitter.
\newblock {On the relativity of inertia: Remarks concerning Einstein's latest
  hypothesis}.
\newblock {\em Proc. Kon. Ned. Acad. Wet.}, 19:1217--1225, 1917.

\bibitem{deSitter2}
W.~de~Sitter.
\newblock {On the curvature of space}.
\newblock {\em Proc. Kon. Ned. Acad. Wet.}, 20:229--243, 1917.

\bibitem{Maldacena}
Juan~Martin Maldacena.
\newblock {The Large N limit of superconformal field theories and
  supergravity}.
\newblock {\em Adv. Theor. Math. Phys.}, 2:231--252, 1998.

\bibitem{Nakahara}
M.~Nakahara.
\newblock {\em {Geometry, topology and physics}}.
\newblock 2003.

\bibitem{GangTian}
Gang Tian.
\newblock {\em {Canonical metrics in Kähler Geometry}}.
\newblock 2000.

\bibitem{NN}
{A. Newlander and L. Nirenberg}.
\newblock {Complex Analytic Coordinates in Almost Complex Manifolds}.
\newblock {\em {Annals of Mathematics}}, {65}({3}):{391--404}, {1957}.

\bibitem{KahlerAdvanced}
Andrei Moroianu.
\newblock {\em {Lectures on Kähler Geometry}}.
\newblock 2007.

\bibitem{comgeo}
Daniel Huybrechts.
\newblock {\em {Complex Geometry: An Introduction}}.
\newblock 2005.

\bibitem{onlykah}
{Yang, Bo and Zheng, Fangyang}.
\newblock {On Curvature Tensors of Hermitian Manifolds}.
\newblock {\em {Communications in Analysis and Geometry}},
  {26}({5}):{1195--1222}, {2018}.

\bibitem{kahcur1}
{Liu, Ke-Feng and Yang, Xiao-Kui}.
\newblock {Geometry of Hermitian manifolds}.
\newblock {\em {International Journal of Mathematics}}, {23}({6}):{1250055},
  {2012}.

\bibitem{kahcur2}
{Liu, Kefeng and Yang, Xiaokui}.
\newblock {Ricci Curvatures on Hermitian manifolds}.
\newblock {\em {Transactions of the American Mathematical Society}},
  {369}({7}):{5157--5196}, {2017}.

\bibitem{diffgeom}
Shoshichi Kobayashi and Katsumi Nomizu.
\newblock {\em {Fundations of Differential Geometry, Vol. 2}}.
\newblock 1969.

\bibitem{complexsitansit}
{Ben-Ahmed, Ali and Zeghib, Abdelghani}.
\newblock {On homogeneous Hermite-Lorentz spaces}.
\newblock {\em {Asian Journal of Mathematics}}, {20}({3}):{531--552}, {2016}.

\bibitem{linalgebra}
Steven Roman.
\newblock {\em {Advanced Linear Algebra}}.
\newblock 2008.

\end{thebibliography}
\end{document}